\documentclass[sigconf]{acmart}
\AtBeginDocument{%
  }

\copyrightyear{2026}
\acmYear{2026}
\setcopyright{cc}
\setcctype{by-nc-nd}
\acmConference[MM '26]{Proceedings of the 34th ACM International Conference on Multimedia}{November 10--14, 2026}{Rio de Janeiro, Brazil}
\acmBooktitle{Proceedings of the 34th ACM International Conference on Multimedia (MM '26), November 10--14, 2026, Rio de Janeiro, Brazil}
\acmDOI{10.1145/3767308.3835653}
\acmISBN{979-8-4007-2213-4/2026/11}
\makeatletter
\renewcommand{\@copyrightpermission}{%
  Copyright held by the owner/author(s). This is the author's version of the
  work. It is posted here for your personal use. Not for redistribution. The
  definitive Version of Record was published in \emph{Proceedings of the 34th
  ACM International Conference on Multimedia (MM '26)},
  \url{https://doi.org/10.1145/3767308.3835653}.}
\makeatother

\newcommand{\numsongs}[0]{9,468}
\newcommand{\duration}[0]{61h}
\newcommand{\hooktheorycnnser}[0]{53.70}

\newcommand{\krn}{\texttt{**kern}}
\usepackage{threeparttable}

\usepackage{array}

\usepackage{tcolorbox}
\tcbuselibrary{listings}
\usepackage{makecell}
\begin{document}

\title{Audio-to-Score Transcription using Pre-trained Features, Data Augmentation, and the New SheetSage-A2S Dataset}


\author{Eoin Cummins}
\authornote{Also with Great Bay University.}
\affiliation{%
  \institution{University College Dublin}
  \city{Dublin}
  \country{Ireland}}

\author{Zhongyi Huang}
\authornotemark[1]
\affiliation{%
  \institution{Guangxi Normal University}
  \city{Guilin}
  \country{China}}

\author{Alexandre D'Hooge}
\affiliation{%
  \institution{Great Bay University}
  \city{Dongguan}
  \state{Guangdong}
  \country{China}}

\author{Zhuoru Mo}
\authornotemark[1]
\affiliation{%
  \institution{Shenzhen University}
  \city{Shenzhen}
  \country{China}}

\author{Yaolong Ju}
\authornote{Corresponding author: juyaolong@gbu.edu.cn}
\affiliation{%
  \institution{Great Bay University}
  \city{Dongguan}
  \state{Guangdong}
  \country{China}}
\email{juyaolong@gbu.edu.cn}


\def\subsectionautorefname{Section}

\providecommand*{\Autoref}[1]{%
  \begingroup
  \def\chapterautorefname{Chapter}%
  \def\sectionautorefname{Section}%
  \def\subsectionautorefname{Section}%
  \def\subsubsectionautorefname{Subsubsection}%
  \def\figureautorefname{Figure}%
  \def\tableautorefname{Table}%
  \def\equationautorefname{Equation}%
  \def\lstnumberautorefname{Line}
  \autoref{#1}%
  \endgroup
}

\begin{abstract}
  Existing audio-to-score (A2S) systems primarily focus on classical music, and the application to popular music remains underexplored. This paper first presents the new SheetSage-A2S Dataset, which includes 61 hours of audio with \texttt{**kern} score encodings for 9,468 clips originating from 6,066 unique songs, the first of its kind to facilitate A2S research for popular music.  Additionally, we improve on existing A2S approaches by using data augmentation and MuQ, a pretrained feature-extraction model for music audio, to enhance generalisation abilities and extract meaningful audio features. Our results show that the proposed A2S model achieves 4.98\% symbol error rate (SER) on the Quartets collection for classical music, which significantly outperforms the 15.3\% SER from the existing state-of-the-art \cite{alfaro-contrerasTransformer2024}. Additionally, our model achieves 20.92\% SER on the SheetSage-A2S dataset for popular music, serving as a strong benchmark for future research. The dataset, model, and code are made publicly available at: \url{https://github.com/Multimodal-Music-Research-Lab/SheetSage2Kern_model}.
\end{abstract}




\maketitle

\begin{figure}[h]
    \centering
    \includegraphics[width=0.95\linewidth]{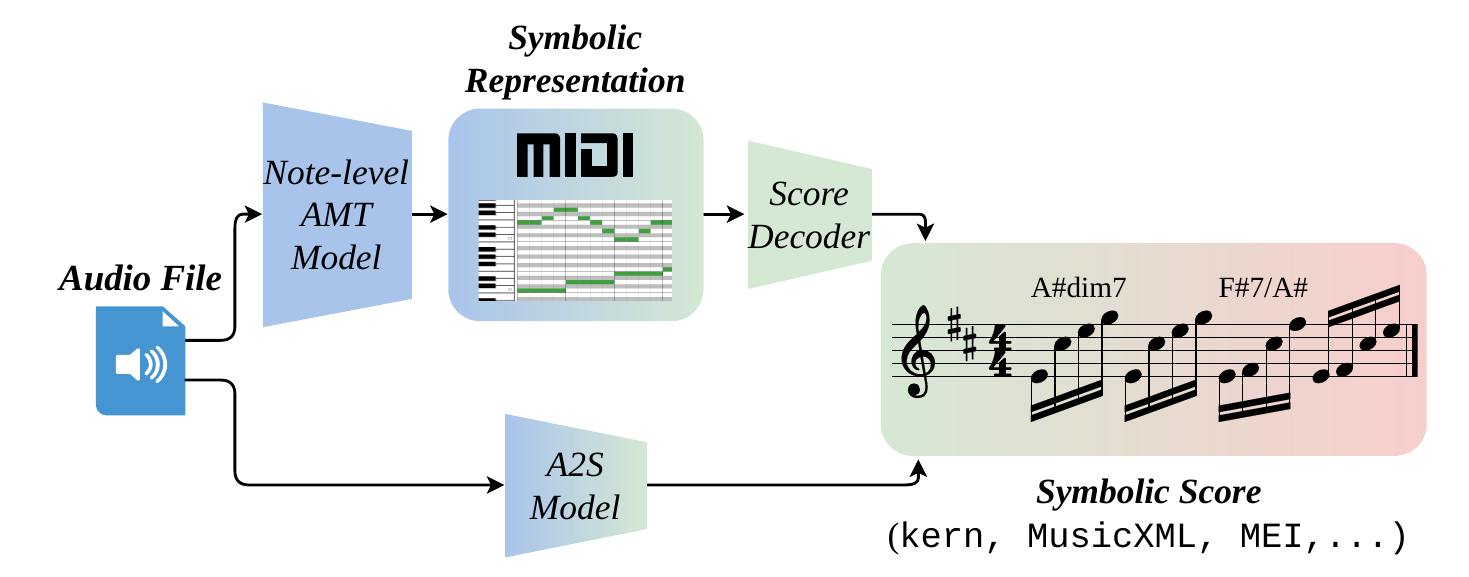}
    \caption{Illustration of the two main ways to generate symbolic music scores: (top) modular approach based on a traditional, note-level AMT model that typically outputs symbolic representations such as MIDI or piano-roll; (bottom) A2S approach that directly generates score symbols based on the input audio file.}
    \Description{A two-part flow diagram. The top path shows an audio file feeding a note-level AMT model, which outputs a symbolic representation shown as a MIDI piano-roll, which then passes through a score decoder to produce a symbolic score. The bottom path shows the same audio file feeding an A2S model that produces the symbolic score directly, without the intermediate representation. The resulting score is a lead sheet excerpt with a melody staff, and chord labels written above it.}
    \label{fig:a2stypes}
\end{figure}
\section{Introduction}\label{sec:intro}

Audio-to-Score Transcription (A2S) can be considered as a subfield of 
Automatic Music Transcription (AMT). Indeed, as defined in \cite{benetosAutomatic2019}, AMT consists in ``the design of computational algorithms to convert acoustic music signals into some form of musical notation''. 
However, more often than not, the output of AMT models is a MIDI or piano-roll representation \cite{benetosAutomatic2019, jamshidiMachine2024}, and not a full-fledged musical score. 
A2S research addresses this limitation, by setting the desired output as a human-readable score.
In this section, we first introduce related work in the field of A2S and limitations of existing approaches, before presenting our proposed approach and contributions.
\begin{table*}[t]
    \caption{Datasets used in A2S research. RISM files are incipits of unknown durations.  All datasets are Western classical music scores except MuseSyn that includes piano arrangements of other musical styles, and our new dataset that gathers lead sheets of popular music. 
    Although our dataset contains audio clips with singing vocals and accompaniment, a polyphonic texture similar to the Quartets, Chorales, and MuseSyn datasets, the corresponding scores of SheetSage-A2S contain the monophonic singing melodies as the essential part of lead sheets.}
    \label{tab:datasets}
    \centering
    \begin{tabular}{m{2cm}cccc>{\centering\arraybackslash}m{4cm}>{\raggedleft\arraybackslash}m{3cm}}
    \toprule
     Dataset    & Num. Files & Duration & Recordings & Score Formats & Music Texture & Instruments
    \\
     \midrule 
     RISM \cite{romanData2020} & 246,870 & Unknown & Synthetic &  PAE \cite{brook1965simplified} incipits & Monophonic & Piano, Harpsichord, Organ, Strings, Winds  \\
     Quartets \cite{romanHOLISTIC2019} & 34,512 & 20.25h & Synthetic & \texttt{**kern} & Polyphonic & String quartets  \\
     Chorales \cite{romanHOLISTIC2019} & 352 & 5.79h & Synthetic & \texttt{**kern}  & Polyphonic & Soprano, Alto, Tenor, Bass voices \\
     MuseSyn \cite{liuJoint2021} & 210 & 9.62h & Synthetic & MusicXML & Polyphonic & Piano \\
     \midrule 
     SheetSage-A2S (Ours) & \numsongs & \duration & Real& \texttt{**kern} & Polyphonic &  Singing vocals for melodies \\
     \bottomrule
    \end{tabular}
\end{table*}
\subsection{Related Work}

Obtaining a formatted score can be done in a modular or end-to-end approach, as illustrated in \autoref{fig:a2stypes}.
Modular approaches involve converting a MIDI or piano-roll intermediate representation to symbolic scores.
This can be done with rule-based algorithms, as in \cite{cogliatiTRANSCRIBING} where the authors compute meter, key signature, and note spellings from MIDI files, and engrave the scores with Lilypond.
MIDI-to-score conversion can also be formulated as a data-driven,  sequence-to-sequence problem, as in \cite{beyerEndtoend2024}, where a transformer model produces MusicXML files from performance MIDI recordings.
A limitation of modular approaches is that they can accumulate errors throughout their processing stages, unlike end-to-end approaches where
audio is directly converted into a typeset score. Convolutional and recurrent neural networks have been used in \cite{carvalhoEndtoend2017, romanEndtoEnd2018, romanHOLISTIC2019, romanData2020} for such end-to-end conversions, as well as attention-based models in \cite{liuAudio2021,alfaro-contrerasTransformer2024}.

The current A2S research faces two major limitations. One is the lack of datasets, which has been highlighted in recent papers
\cite{romanData2020, arroyoNeural2022}, as many datasets include either audio files or symbolic scores but rarely both.\footnote{See for instance \url{https://www.audiocontentanalysis.org/datasets} (all web links were last accessed and verified in March 2026).} One way of circumventing this problem is to generate audio files from symbolic scores using audio synthesis with virtual instrument sounds. 
The \texttt{humdrum-data} repository\footnote{\url{https://github.com/humdrum-tools/humdrum-data}} is therefore sometimes used as a source of music scores in the \texttt{**kern} format \cite{sappOnline2005} to build synthetic A2S datasets such as \textit{Quartets} and \textit{Chorales}, used in \cite{alfaro-contrerasTransformer2024, arroyoNeural2022, romanHOLISTIC2019}. 
Other public music scores websites like the RISM and Musescore catalogues are used in \cite{romanData2020, liuJoint2021}, but the audio files are still obtained through digital synthesis.
While synthesis facilitates dataset creation, resorting to such recordings is a significant drawback that questions the ability of previous work to generalise to real recordings. Furthermore, all the aforementioned papers study Western Classical Music, leaving the vast genre of Popular Music mostly untouched. 
\autoref{tab:datasets} summarises the characteristics of datasets used in previous A2S research.

The other limitation is the lack of exploiting popular machine learning techniques to improve performance, such as the use of pre-trained models and data augmentation.
Pre-trained models proved useful in AMT, as \cite{rileyHigh2024} finetune a model trained on piano music and attain state-of-the-art performance on guitar music. The student-teacher learning paradigm
can also be used and showed promising results in drums transcription \cite{wuAutomatic2017}.
Recently, new music foundation models have also been proposed, which offer learned representations that can be applied to downstream tasks \cite{maFoundation2024}. 
Choosing the right pre-trained model for a downstream task is not straightforward, which is why benchmarks like MARBLE \cite{yuanMARBLE} are useful to assess the performance of a model.
This benchmark uses 12 datasets to evaluate models on 18 musical understanding tasks like key detection, chord estimation or source separation.
Examples of models performing well on MARBLE are MERT \cite{liMERT2024}, MusicFM \cite{wonFoundation2024}, and MuQ \cite{zhuMuQ2025}. All three models 
are trained using masked language modelling, but with different tokenisation strategies. 
Using a music foundation model is a promising way of improving A2S models, as transcribing recordings to human-readable scores involves several complex processes, such as pitch and chord recognition, beat detection, and key finding.
Regarding data augmentation, common techniques consist of applying controlled pitch-shifting or time-stretching to audio recordings \cite{mcfeeSOFTWARE}. 
These augmentations were shown to increase performance in AMT-related tasks, including transcriptions of lyrics \cite{zhangPDAugment2021}, drums \cite{jacquesData2019}, piano \cite{edwardsDataDriven2024}, or classical music pieces with varied instrumentations \cite{8461686}. 
Nevertheless, we could not find A2S papers that use either pre-trained models or such data augmentation techniques.
\begin{figure*}
    \centering
    \includegraphics[width=0.95\textwidth]{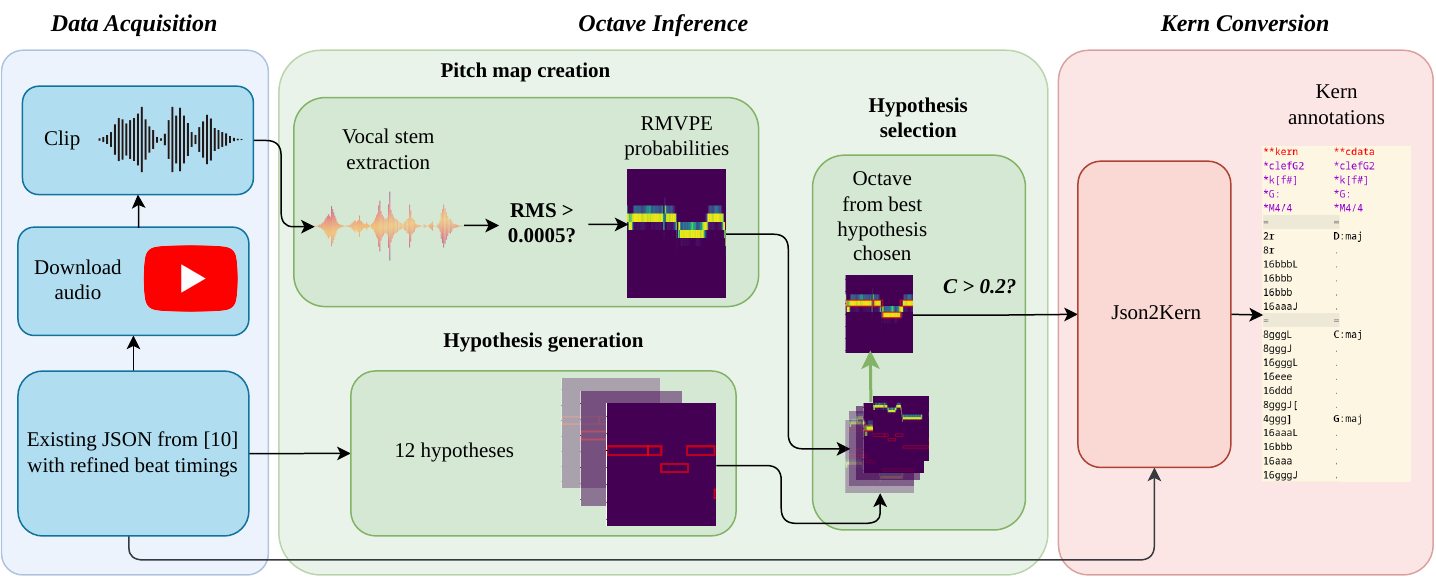}
    \caption{Overview of the SheetSage-A2S dataset creation pipeline, with three main steps that are introduced in detail at \autoref{sec:dataset}. The confidence score $C$
    measures how well a potential octave and starting beat align with RMVPE's \cite{wei_rmvpe_2023} vocal pitch estimates. The process of Json2Kern is further illustrated in \autoref{fig:json2kern}.}
    \Description{A block diagram showing the three main stages of the dataset creation pipeline flowing from left to right. It starts with a blue Data Acquisition section where existing JSON files are used to download YouTube audio, resulting in an audio clip. The middle green section, Octave Inference, takes that audio clip through vocal stem extraction. If the RMS energy is above 0.0005, it generates RMVPE pitch probabilities. Meanwhile, the original JSON generates 12 hypotheses. A hypothesis selection step compares the confidence scores of 12 hypotheses to find the most likely octave. If the confidence score C is greater than 0.2, the data moves to the final red section called \texttt{**kern} Conversion. Here, the chosen octave and the original JSON data are fed into a Json2Kern module, which outputs the final \texttt{**kern} annotations.}
    \label{fig:dataset}
\end{figure*}
\subsection{Contributions}

In this paper, we address the data scarcity limitation by proposing the new SheetSage-A2S dataset, which includes 61 hours of audio with \numsongs~ \texttt{**kern} score encodings for 6,066 unique songs,\footnote{Each encoding represents a section (e.g. chorus, verse) of the song. Therefore, multiple encodings may belong to the same song.} the first of its kind to facilitate A2S research for popular music. The scores in SheetSage-A2S are \textit{lead sheets}, i.e. scores that engrave the melody and chords for popular music and jazz songs \cite{martinez-sevillaOptical2025, donahue2022melody}. A lead sheet excerpt is shown on the right of  \autoref{fig:a2stypes}, where the melody is specified in the staff while chords are attached above the melody. Unlike existing A2S datasets, all music from our SheetSage-A2S dataset contains real recordings released commercially that can be found on YouTube, or other music streaming services. Another difference is that our dataset isolates the melody tracks, which are all sung by singers, and exclusively records their information from polyphonic mixtures, in contrast to the existing databases that transcribe and notate every individual note of relevant musical instruments. 

We also address the performance ceiling by proposing our new A2S model, which adopts an autoregressive transformer decoder similar to \cite{alfaro-contrerasTransformer2024}, and make three modifications (shown in \Autoref{fig:muq}). First, we increase the capacity and stability of the original decoder by expanding its feedforward dimension from 256 to 1024 and adopting a Pre-Norm architecture. Second, we replace the vanilla CNN encoder with MuQ \cite{zhuMuQ2025}, a pre-trained model that employs self-supervised learning to extract rich audio features for music understanding. Finally, we employ data augmentation, including pitch shifting and time stretching, to improve robustness and increase the amount of data for training. 

The experimental results demonstrate that the proposed A2S model achieves significant performance improvements over the current state-of-the-art method \cite{alfaro-contrerasTransformer2024} on the Quartets dataset for classical music. Meanwhile, it obtains a Symbol Error Rate (SER) of 20.92\% on the new SheetSage-A2S dataset for popular music, establishing a strong benchmark for future research in this domain.

The organisation of the rest of this paper is described hereafter. In \Autoref{sec:dataset}, we present how the new SheetSage-A2S dataset was pre-processed and retrieved. \Autoref{sec:methodology} describes the baseline CNN model used in this study and our newly proposed architecture, which are compared through extensive experiments presented in \Autoref{sec:experiments}. \Autoref{sec:results} details the transcription performance results and ablation studies. In \Autoref{sec:dis}, discussions are made based on the experimental results, with inference results to further illustrate the model's performance. Finally, we conclude our work and propose potential directions for future research in \Autoref{sec:future}.

\section{Dataset}
\label{sec:dataset}

We introduce SheetSage-A2S, a dataset for popular music with audio-to-score transcription built from the original SheetSage annotations \cite{donahue2022melody} and corresponding audio. At SheetSage, each annotation represents a distinct segment (such as a verse or chorus) from a particular song. We therefore refer to these individual segments as ``clips''.  The dataset contains \numsongs\ clips derived from 6,066 songs by 2,891 artists. The mean clip duration is approximately 23 seconds, yielding 61 hours of audio in total. Each example pairs an audio clip with a \texttt{**kern} score obtained from the source annotations, and we choose \texttt{**kern} as the score encoding format due to its simple and compact form, compared to other formats such as MusicXML.\footnote{\url{https://www.musicxml.com/}}
Our dataset release contains no copyrighted audio: we distribute the \texttt{**kern} annotations under CC-BY-NC-ND 4.0 (similar to the original SheetSage), along with YouTube links, timestamps, and durations. We also provide pre-computed audio features for all clips.
The whole dataset creation pipeline for SheetSage-A2S is shown in \Autoref{fig:dataset}, where we first acquire the data from SheetSage (\autoref{sec:scrape}), then we clean the data by removing all the clips without melodies or chords. For the remaining clips, we obtain the pitches for all singing melodies via the process of octave inference (\autoref{sec:data_cleaning}). Finally, we compile all relevant music elements for each clip from the JSON file and convert them into the resulting \krn~annotations (\autoref{sec:json2kern}), which are combined with the downloaded audio to form the proposed SheetSage-A2S dataset. 
 
\subsection{Data Acquisition}\label{sec:scrape}

We follow \cite{donahue2022melody} and use their open-source JSON file,\footnote{\url{https://github.com/chrisdonahue/sheetsage-data/blob/main/hooktheory/Hooktheory.json.gz}} which contains clips of annotations as well as the corresponding YouTube links for music audio, as a starting point. We attempt to scrape all the audio files with YouTube links using the \textit{yt-dlp}\footnote{\url{https://github.com/yt-dlp/yt-dlp}} 
toolkit. 
There are 26,175 entries in the original JSON file, and we managed to download 20,072 audio clips. Among the 6,103 failed cases, around 4000 clips are not available anymore, the remaining ones are not made public due to copyright reasons. 

Within the JSON file provided by \cite{donahue2022melody}, the onsets and offsets for each note and chord are specified in terms of musical beats rather than exact timestamps in seconds. To align these symbolic annotations with the downloaded audio, the JSON includes refined beat-to-time mappings generated using the beat and downbeat detection algorithms \cite{bockJoint2016} from the \textit{madmom} library \cite{bock2016madmom}, which has been demonstrated to be highly reliable for mapping beat positions to precise timestamps in seconds. We adopt the formalisation from \cite{donahue2022melody} and define this function as an alignment function $A : [0, B) \to [0, T)$  that assigns each of  $B$  beats in the metrical structure to a time  $t \in [0, T)$  in the audio.

\subsection{Octave Inference}
\label{sec:data_cleaning}

One obstacle encountered during the step of data acquisition is that the melodies written on the original SheetSage annotation use a relative octave system rather than an absolute one.\footnote{This means that the annotation only contains the pitch class and its octave relative to other notes.} As a result, the exact pitch of each note from the melody is unknown, preventing us from obtaining symbolic scores.

\textbf{Pitch map creation.}  To overcome this issue and obtain absolute octave information, we use existing pitch estimation tools to obtain fundamental frequencies (F0) for singing melodies,\footnote{We choose to focus singing melodies rather than melodies performed by instruments in this work, because existing pitch estimation tools \cite{kim_crepe_2018, wei_harmof0_2022, wei_rmvpe_2023} only used synthesised instruments for testing, which may not generalise well on our dataset that contain real recordings for commercial music.} which are first extracted using a
source separation tool known as the Mel-band Roformer \cite{wangMelRoFormer2024} for each audio clip. We then eliminate all silent clips and retain the ones whose singing RMS (Root Mean Square) energy exceeds 0.0005, resulting in 16,310 candidate clips. Next, we adopt RMVPE \cite{wei_rmvpe_2023} as our pitch estimation tool and apply it to the isolated singing melodies,  resulting in
72 semitone-spaced bins
that form a pitch probability  matrix $P\in [0,1]^{T \times72}$, where T denotes the number of time frames and the y-axis contains semitone-spaced frequency probabilities from the note range within C1-B6.

\begin{figure}[t]
    \centering
    \includegraphics[width=0.7\linewidth, trim={1cm 1cm 2cm 0}]{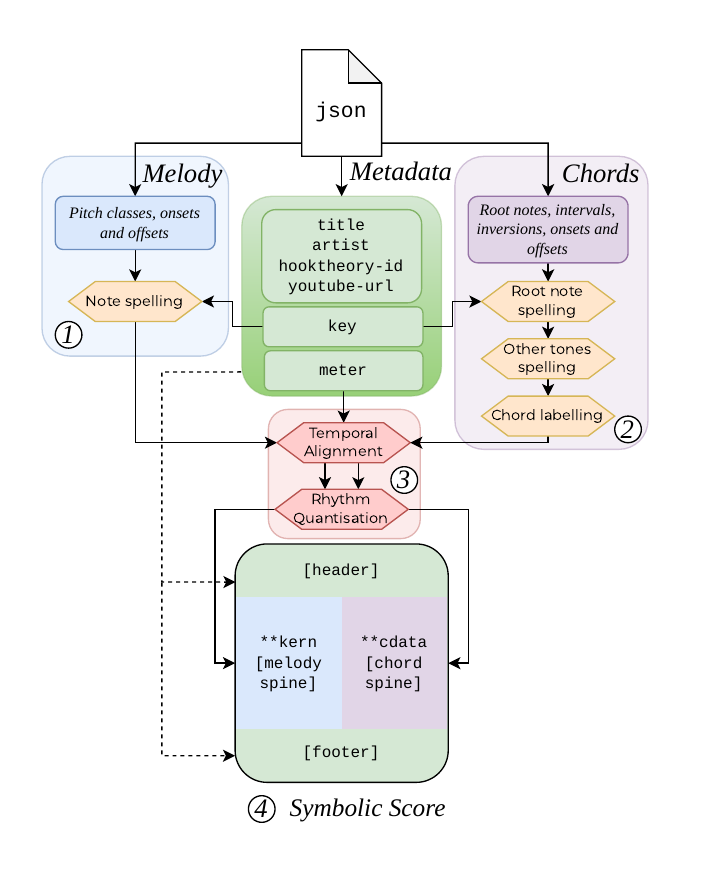}
    \caption{Summary of the json to \texttt{**kern} conversion pipeline. Numbers denote the four main processing steps described in \autoref{sec:json2kern}.}
    \label{fig:json2kern}
\end{figure}

\textbf{Hypothesis generation.} Let $N$ denote the set of  annotated 
notes in the annotated melody, where each note $n \in N$ has a relative pitch $p_{rel,n}$ and beat-indexed onset and offset $[b_{start,n}, b_{end,n}]$. Each relative pitch $p_{rel,n}$ is made up of the note's pitch class (an integer from 0 to 11) and the note's relative octave (an integer typically between -1 and 1, which allows users to enter for example an E and then an E the next octave up or down). To determine the most probable octave, we first treat the note's relative pitch as an absolute pitch. We then shift it upwards by an octave for all octaves within RMVPE's output range. We test all combinations of six octave shifts $O \in \{0,1,2,3,4,5\}$ and two start time offsets $s$: the \textit{madmom} detected downbeat start timestamp and the user-defined start timestamp. Since the alignment function $A$ maps beats to global time, but our pitch matrix $P$ is for a trimmed clip, we subtract $s$ to find the notes' local positions. We test both offsets because \textit{madmom} can sometimes miss the first beat (e.g. if a clip begins with a held note), which shifts the entire melody.
For each note we then calculate the start and end time as: 

$$t_{start,n}(s) = A(b_{start,n}) - s, \quad t_{end,n}(s) = A(b_{end,n}) - s$$

Each combination of $(O, s)$ defines a hypothesis index mask: 
 the set of cells in $P$ that the notes would occupy under that 
hypothesis, shown as the red rectangles in 
Figure~\ref{fig:dataset}:

$$H(O, s) = \bigcup_{n \in N} \bigl\{(t,\ p_{rel,n} + 12 \cdot O) \mid t \in [t_{start,n}(s),\ t_{end,n}(s)]\bigr\}$$

Combining six octave shifts with two start time offsets gives 12 
hypotheses per clip.

\textbf{Hypothesis selection.} We score each hypothesis by the 
probability mass from $P$ that falls inside its mask, normalised by 
total note duration:

$$C(H) = \frac{\sum_{(t,\,p)\,\in\, H} P(t,\, p)}{\sum_{n \in N}\bigl(t_{end,n}(s) - t_{start,n}(s)\bigr)}$$

The starting time and octave from the hypothesis with the highest confidence are selected as the clip’s octave and starting time. We discard clips with confidence $C = \underset{H}{\mathrm{max}}~C(H)$ below a threshold of 0.2 for the following reasons. First, manual inspection revealed that confidence scores below 0.2 indicated severe transcription errors or instances where the instrumental part, rather than singing melodies, was transcribed. 
Second, through a manual inspection of around 150 clips as a quality check, we did not find any that were assigned the incorrect octave with confidence above 0.2. As a result, we obtain the clips with only singing vocals with correct absolute octave information, resulting in \numsongs\ clips that are ready to be converted as \texttt{**kern} scores in the following step.

\subsection{\texttt{**kern} Conversion}
\label{sec:json2kern}

The original SheetSage files are in JSON format. 
To use the files in our A2S pipeline, we convert them to the \texttt{**kern}\footnote{\url{https://www.humdrum.org/rep/kern/}} format. 
This conversion process is illustrated in \autoref{fig:json2kern} and based upon four main steps: converting the melody notes, converting the chord information, aligning and quantising the rhythm of the melody and chords, and finally merging everything in a Humdrum \krn~file.

\textit{(1) Melody: } 
The original JSON file represents the pitch class of each melodic note using an integer between 0 and 11, but this representation is enharmonic\footnote{In music theory, enharmonic refers to two or more distinct note spellings (e.g. note names with different accidentals) that correspond to the same exact pitch class in twelve-tone equal temperament.
For example, C$\sharp$ and D$\flat$ are enharmonic equivalents: they sound identical, but are written differently in the score and serve different harmonic and contextual functions in a piece of music.} and does not specify note spelling, which is essential for standard music score generation. From the perspective of music theory, note spellings should be consistent with the underlying key signature \cite{rushtonEnharmonic2001}.
We therefore extract the key signature information from the source JSON files, 
and derive the note spellings from pitch classes accordingly.

\textit{(2) Chords: }
Since the SheetSage-A2S dataset represents lead sheets, chords are annotated in addition to the melody. Chords are defined by root pitch class, interval content in semitones, and inversion. \textit{Root note spelling} is obtained in the same way as melody notes if it belongs to the scale. For out-of-scale notes, enharmonic spelling is chosen by matching the key's alteration type (e.g. using sharps in a key with sharps like D Major). \textit{Other tones spelling} is conducted with respect to the root note, matching the intervallic content of the chord to a dictionary of known chord natures to choose the correct enharmonic spelling.
The final \textit{chord labelling} step produces a label in Harte syntax \cite{harteSymbolic2005} of the form \texttt{<root>:<nature>[/bass]}, such as  \texttt{A:min7/3} for an A minor seventh chord in first inversion, for example. 

The \texttt{<bass>} component of the chord label is omitted if it is identical to the root note, and the \texttt{<nature>} is obtained from matching the intervallic content to a pre-existing chord database that was assembled as needed while preparing the dataset.

\textit{(3) Temporal Alignment and Rhythmic Quantisation:}
The melody notes and chord annotations are encoded independently, each 
carrying its own beat-indexed onset and offset values. 
For \textit{temporal alignment}, we merge all note and chord 
events, then iterate over the resulting time slices. At each slice, we build the \texttt{**kern} and \texttt{**cdata} spines by writing the melody token and the current chord symbol as a 
tab-separated row, with a dot (\texttt{.}) filling any column where 
no new event begins.
We then conduct \textit{rhythm quantisation} by mapping the duration values to \texttt{**kern} rhythm tokens, using rational fractions with a maximum denominator of 96 to deal with fast grace notes. For example, durations of $1$ and $0.5$ (a quarter note and an eighth note) in the JSON file becomes respectively $4$ and $8$ in \krn.

\textit{(4) Symbolic Score:}
We finish the conversion by merging the two musical spines \texttt{**kern} (for the melody) and \texttt{**cdata} (for the chords) with a header and a footer.
The header declares global attributes including clef, key and time 
signatures, as well as artist and clip metadata.
The footer appends bibliographical records including the original 
HookTheory\footnote{\url{https://www.hooktheory.com/theorytab}, HookTheory is the Internet’s largest database of songs broken down by music theory, which is also the source of the original SheetSage annotations.} and YouTube links.

\subsection{Data Statistics} 

\begin{figure}
    \centering
    \includegraphics[width=\linewidth]{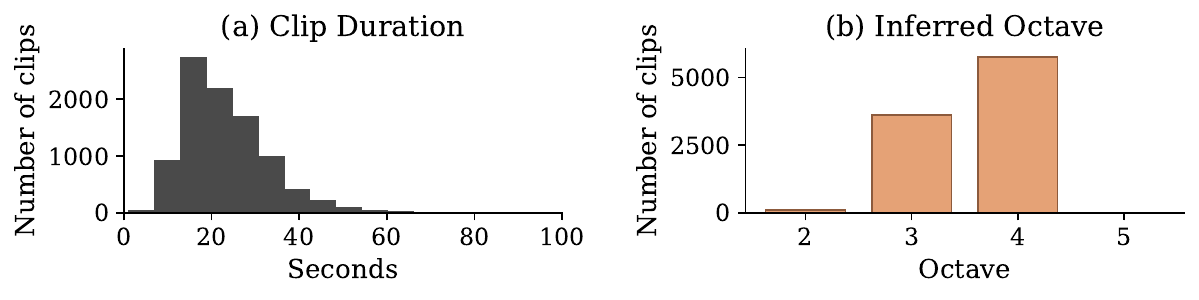}
    \caption{SheetSage-A2S dataset statistics. (a) Clip duration, truncated at 100 seconds; (b) Inferred absolute octave.}
    \label{fig:dataset_statistics}
\end{figure}
The \numsongs\ clips (\autoref{fig:dataset_statistics}) have a mean duration of 23.36 seconds with a standard deviation of $\sim$11.34 seconds; 12 clips exceed 100 seconds.
The clips originate from 6,066 unique songs across 2,891 artists. While 3,635 songs are represented by a single clip, the remainder contribute several. The songs per artist distribution is long-tailed: 2,005 artists contribute only one song, with an overall mean of 2.1 songs per artist. A notable outlier is The Beatles, with 127 songs. 
Using the octave inference process introduced in \Autoref{sec:data_cleaning}, 98.66\% of clips are assigned to octave three or four. 

\section{Proposed Method}

\label{sec:methodology}

This section describes our proposed A2S method. We first detail the core model architecture, including the baseline design and our enhancements leveraging a pre-trained audio foundation model. We then discuss the 
data augmentation techniques implemented.

\subsection{Model Configuration}\label{sec:config}

\begin{figure}
    \centering
    \includegraphics[width=0.95\linewidth]{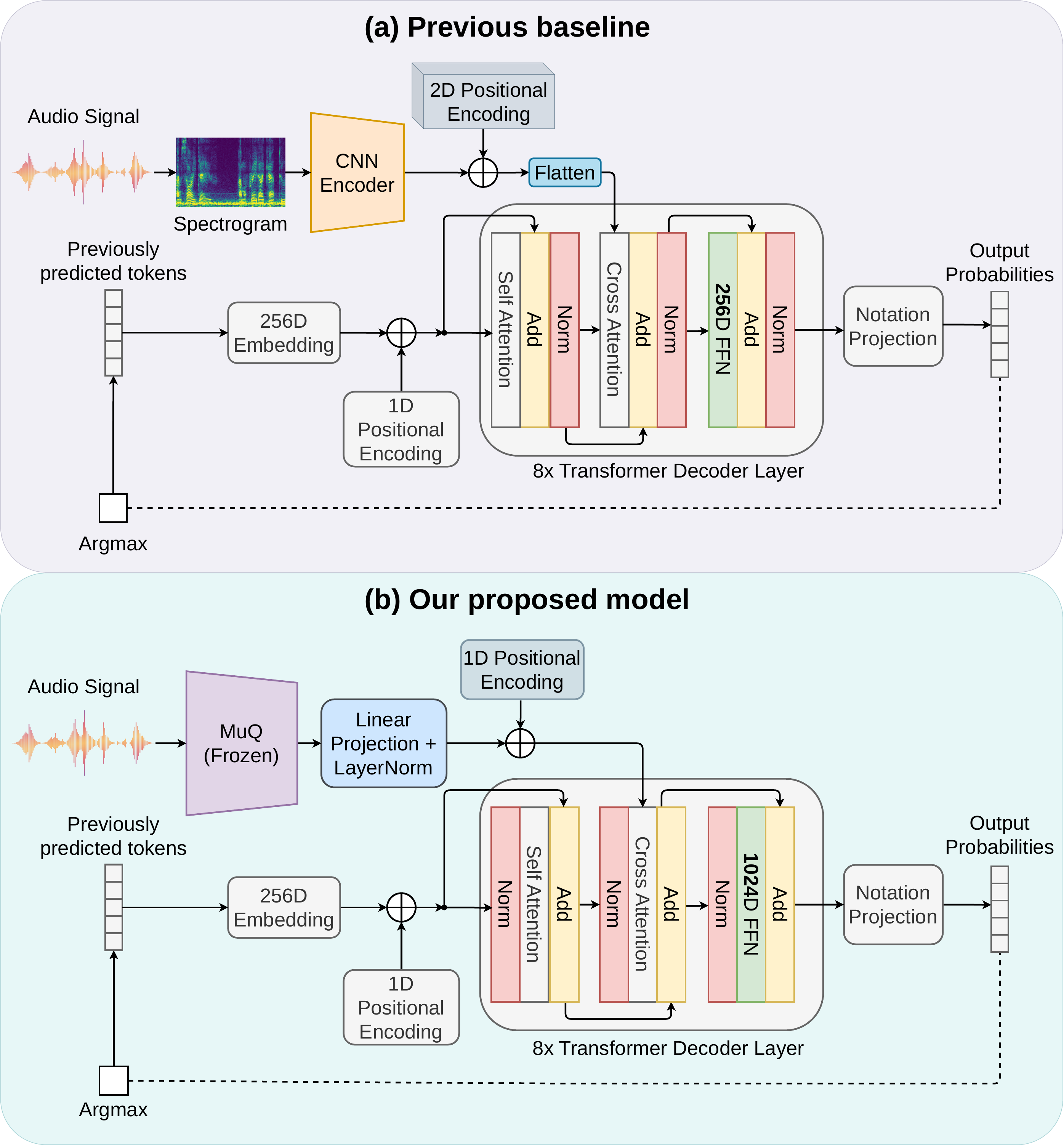}
     \caption{Comparison of our proposed architecture against the baseline of \cite{alfaro-contrerasTransformer2024}.   Both models use an autoregressive Transformer decoder to generate \texttt{**kern} score symbols. We replace the CNN encoder from the baseline with MuQ \cite{zhuMuQ2025}, a pre-trained model to extract audio features, followed by a linear projection and LayerNorm to map MuQ's embedding dimension from 1024 to 256. We also increase the decoder feedforward dimension from 256 to 1024, and change its normalisation order from post-norm to pre-norm. }
     \Description{A two-part diagram comparing the previous baseline architecture and the proposed model. The top panel shows the baseline, which processes an audio spectrogram using a CNN encoder with 2D positional encoding. This feeds into an 8-layer Transformer decoder that uses post-normalization and a 256-dimensional feedforward network. The bottom panel displays the proposed model, which replaces the CNN with a frozen MuQ model processing the audio signal, followed by a linear projection, layer normalization, and 1D positional encoding. The proposed model's decoder also uses 8 layers but shifts to a pre-normalization structure and expands the feedforward network to 1024 dimensions.}
    \label{fig:muq}
\end{figure}

\textit{Baseline Architecture.}
We start with the model from \cite{alfaro-contrerasTransformer2024} shown in \autoref{fig:muq}(a). This model is a CNN encoder with a Transformer decoder. The CNN encoder first turns an input spectrogram into a 2D feature map. After adding 2D positional encodings, the feature map is flattened to form the cross-attention memory passed to the Transformer decoder. 

\textit{Proposed Architecture.}
Our newly-proposed A2S model is illustrated in \autoref{fig:muq}(b). To establish a stronger foundation, we first increase the capacity of the Transformer decoder by expanding the Transformer decoder's feedforward dimension from 256 to 1024 and changing its normalisation order from post-norm to pre-norm in order to improve training stability \cite{xiong2020layer}. We refer to this change as the \textit{1024-PreNorm} baseline, which will be further studied in \Autoref{sec:performance}. Then, we adopt the MuQ pretrained model \cite{zhuMuQ2025} to replace the vanilla CNN encoder as a feature extractor from audio files, to study whether it can mitigate data scarcity issues discussed in \Autoref{sec:intro}. We utilise MuQ due to its strong overall performance on the music understanding benchmark MARBLE \cite{yuanMARBLE}, suggesting that it learns useful representations for downstream music tasks.

We specially keep MuQ frozen and use its final layer's hidden-state sequence as output. We then use a linear projection followed by LayerNorm to reduce the MuQ embedding size from its original 1024 to 256. 

As MuQ outputs a one-dimensional sequence, we use a one-dimensional sinusoidal positional encoding in place of the baseline's 2D one. Finally, the resulting features (note that MuQ outputs 25 feature vectors per second, compared to $\sim$5.38 per second from the baseline CNN encoder) are then input as the cross-attention memory to the Transformer decoder.

\subsection{Data Augmentation}

To improve robustness and increase the amount of training data, we apply offline data augmentation to the training split. We implement pitch-shifting and time-stretching, leaving the exploration of other augmentation techniques to future work.
We denote pitch shifting by $\mathcal{P}_k$, where $k$ is measured in semitones, and time stretching by $\mathcal{T}_s$, where $s$ is the stretch factor. For each training pair $(x,y)$, we generate six augmented variants, one for each pitch shift $k \in \{-3,-2,-1,1,2,3\}$. For each such variant, we independently sample a time-stretch factor $s \in \{0.9, 0.95, 1.05, 1.1\}$ and construct augmented audio $x' = \mathcal{T}_s(\mathcal{P}_k(x))$. These ranges were chosen to provide the model with enough variety to learn to handle different frequency ranges and tempos while ensuring the augmented audio still sounds like a real performance.

The symbolic target is transposed accordingly, giving $y'=\mathcal{P}_k(y)$. We do not modify $y$ for $\mathcal{T}_s$ on the SheetSage-A2S dataset. As the target representation uses relative note values and lacks an explicit meter, it is tempo-invariant. However, for the Quartets dataset, the tempo metadata is updated accordingly. The original training example is retained alongside its augmented variants, resulting in a $7\times$ expansion of the training set. We use Rubber Band\footnote{\url{https://breakfastquay.com/rubberband/}} for both pitch shifting and time stretching. 

\section{Experiments}
\label{sec:experiments}

We conduct experiments on two datasets: the Quartets collection and our new SheetSage-A2S dataset. We provide implementation details for training and testing the baseline model and our proposed architecture.

\subsection{Configurations}

\textbf{Quartets Dataset.}
We evaluate our approach on the Quartets collection, which comprises string quartets by Mozart, Beethoven, and Haydn, serving as a direct point of comparison with \cite{alfaro-contrerasTransformer2024}. The dataset is partitioned into 70\% training, 15\% validation, and 15\% test sets, consistent with \cite{alfaro-contrerasTransformer2024}.
For our experiments on this dataset, we train a baseline A2S model using the original word-level tokenisation. We then incrementally incorporate MuQ, the pre-trained encoder and our data augmentation strategy to evaluate their individual contributions. 

\noindent
\textbf{SheetSage-A2S Dataset.}
 Following the methodology in \cite{donahue2022melody}, this dataset employs an artist-stratified split of 80\% training, 10\% validation, and 10\% testing.  During our data analysis in \autoref{sec:data_cleaning}, we observed that octave detection confidence scores served as a proxy for overall transcription quality. While we used a cut-off of 0.2 for the dataset, we applied a stricter threshold of 0.45 exclusively to the validation and test sets. This strategy ensures that our models are evaluated against accurate ground truths. We use word-level tokenisation, following the results of our tokenisation ablation in \autoref{sec:token_ablation}.

For both datasets, the audio files were resampled to 22,050 Hz when using the baseline architecture, following \cite{alfaro-contrerasTransformer2024}, and to 24,000 Hz when using MuQ \cite{zhuMuQ2025}. In both cases, we first train a baseline A2S model. We then evaluate the incremental improvements from incorporating MuQ and our data augmentation strategy.

\subsection{Implementation Details}

\textbf{Evaluation.}
Symbol Error Rate (SER) computes the Levenshtein distance between ground truth and prediction: the minimum number of single-element edits (insertions, deletions, or substitutions) required to transform the predicted sequence into the target sequence. We use SER as our primary metric for both datasets, with all sequences generated using greedy decoding at inference time. This ensures direct compatibility with existing A2S work \cite{alfaro-contrerasTransformer2024}, and also provides a consistent baseline against which future models trained on SheetSage-A2S can be evaluated. While MV2H \cite{mcleod2018evaluating} has been used in other audio-to-score work, current implementations do not support the lead sheet format we generated in \autoref{sec:json2kern}. We therefore provide both a per-spine SER breakdown and SER scores for both melody and chord instead.

\noindent
\textbf{Training.}
Models are trained using the AdamW optimiser with a weight decay of 1e-3, a constant learning rate of 5e-5, and a batch size of 8. This contrasts with the original training setup (Adam, no weight decay, lr 1e-4, batch size 1). 
We made this switch as we found through empirical testing that it improved training stability. We use an early-stopping patience of 5 epochs on the validation loss with a minimum delta of 1e-3, and retain the best model.

On both datasets the model is trained with cross-entropy loss. As the SheetSage-A2S annotations are user-generated and contain occasional errors, we apply label smoothing of 0.1 while training on SheetSage-A2S, which softens the target distribution and reduces the penalty for deviating from potentially incorrect ground truth labels. We found this label smoothing provided a small SER decrease. 

\section{Results}
\label{sec:results}

\subsection{Transcription Performance}\label{sec:performance}
The transcription performance on both datasets is detailed in \autoref{tab:results}, using the word-level tokenisation scheme identical to the baseline model \cite{alfaro-contrerasTransformer2024}. On Quartets, our proposed model achieves an SER of 4.98, representing a 67.5\% relative reduction in error compared to the previous baseline of 15.3. Performance on the newly introduced SheetSage-A2S dataset demonstrates the inherent challenges of popular music transcription for lead sheets, with an SER of 20.92 for our proposed model. 
For a more granular and musically meaningful evaluation, we also include the SER per spine, namely melody and chord SER results for our SheetSage-A2S dataset in \autoref{tab:results}. Every proposed modification improves all of these scores.  Note that per-spine SER is computed excluding structural tokens (such as tab and newline), which are relatively easier to predict, so per-spine SER values are expected to be higher than normal SER. SER can also exceed 100\%, typically when the model produces many spurious insertions. 

\begin{table}[t]
\centering
\small
\caption{Symbol error rate (SER) for each proposed modification, with a granular breakdown (melody and chord SERs) on SheetSage-A2S. Each row adds one component to the baseline of \cite{alfaro-contrerasTransformer2024}. The spine-level metrics exclude structural tokens (tabs and newlines), which are comparatively easy to predict, and are therefore expected to be higher than the overall SER. Melody and chord SERs are not provided for the Quartets dataset, as it doesn't contain chord annotations, and melodic content can span multiple distinct spines. }
\label{tab:results}
\setlength{\tabcolsep}{5pt}
\begin{tabular}{lccccc@{\hskip 12pt}ccc}
\toprule
& \multicolumn{3}{c}{\textbf{SheetSage-A2S}} & \multicolumn{1}{c}{\textbf{Quartets}} \\

\textbf{Variant} & \textbf{SER} & \textbf{Melody SER} & \textbf{Chord SER} & \textbf{SER} \\
\midrule
Baseline \cite{alfaro-contrerasTransformer2024} & 66.85 & 106.88 & 64.12 & 15.3 \\
+1024-Pre-Norm & 53.70 & 90.56 & 51.80 & 8.48 \\
+MuQ encoder & 25.39 & 46.21 & 26.76 & 7.16 \\
+Augmented data & \textbf{20.92} & \textbf{38.62} & \textbf{22.28} & \textbf{4.98} \\
\bottomrule
\end{tabular}
\end{table}

\subsection{Ablation study}
\label{sec:token_ablation}

\textit{Encoder and augmentation.} 
To evaluate the individual contributions of the proposed methods, in \autoref{tab:results} we analyse the performance of intermediate models bridging the  baseline \cite{alfaro-contrerasTransformer2024} and the final architecture. The evaluation isolates three primary modifications: the model architecture finetuning, the integration of the MuQ feature extractor, and the application of data augmentation. 

First, our 1024-PreNorm baseline (see \Autoref{sec:config}) achieves an SER of 8.48 on the Quartets dataset and \hooktheorycnnser\ on SheetSage-A2S. Based on our preliminary experiments, we primarily attribute the  improvement of the original model's 15.3 SER \cite{alfaro-contrerasTransformer2024} on Quartets and 66.85 on SheetSage-A2S to the increased feedforward dimension from 256 to 1024. Next, replacing the vanilla CNN encoder with frozen MuQ features further reduces the Quartets SER to 7.16. More notably, for SheetSage-A2S, adding MuQ decreases the SER to 25.39. The larger performance gain on SheetSage-A2S is expected, as the MuQ model we utilise was pre-trained on the Million Song Dataset (MSD), a large corpus of popular music. While songs in SheetSage-A2S may overlap with the MSD, quantifying this is non-trivial due to naming inconsistencies across the datasets. However, the consistent gains from MuQ on the stylistically distinct Quartets corpus suggest the improvements stem from generally useful learned representations. Finally, applying data augmentation yields our best results: a 4.98 SER on Quartets and 20.92 SER on SheetSage-A2S. 

\textit{Tokenisation.} To determine the most effective tokenisation strategy, we conducted an ablation study comparing the three methods proposed by \cite{martinez-sevillaOptical2025}: word-level, mid-level, and character-level tokenisation. All experiments were run using our MuQ architecture prior to applying data augmentation.
As shown in \autoref{tab:tokenisation_results}, word-level tokenisation outperformed both alternatives. Although mid-level tokenisation performed best in \cite{martinez-sevillaOptical2025} and offers theoretical advantages such as a bijective mapping between graphical symbols and tokens, and a reduced vocabulary size, these advantages did not translate to our dataset. We partly attribute this to the long sequence lengths present in the SheetSage-A2S dataset, where the reduced sequence length produced by word-level tokenisation may be more beneficial.

\begin{table}[tbhp] 
\centering
\caption{Performances of different tokenisation methods on the SheetSage-A2S dataset. The model used is the MuQ variant without augmentation.} 
\label{tab:tokenisation_results}
\begin{tabular}{lccc}
\hline
Method & Symbol Error Rate  & Token count\\
\hline
Character level & 33.97 &  61\\
Mid level & 30.44 & 224\\
Word level & \textbf{25.39} & \textbf{4763}\\
\hline
\end{tabular}
\end{table}

\section{Discussion}\label{sec:dis}

In this section, we discuss some limitations observed from using the SheetSage-A2S dataset, in particular due to the non-standardised nature of the annotations. 
We then propose a qualitative analysis of a transcription result from our best performing model to reflect upon the strengths and weaknesses of our approach.

\subsection{User-Annotation Variabilities}

A contributing factor to the higher SER on the SheetSage-A2S dataset is the inherent variability of user-generated content. Transcribers slightly differ in subtle stylistic choices, which the model has no way of discerning from the audio. We choose to release the core dataset with minimal preprocessing. However, this lack of standardisation is likely to introduce performance penalties.

First, users transcribe melodies and chords with varying standards. Some annotators label a chord only once when the harmony changes, while others annotate the same chord repeatedly for every rhythmic strum, or once per bar. 
Furthermore, we observe issues with inconsistent transcriptions, where some users stop transcribing before the audio clip officially ends as the remaining (usually small) part does not belong to the current music section, which penalises the model for correctly transcribing the remaining audio. 
Some users also transcribe small parts of the accompaniment when vocals are silent, a behaviour our model did not catch on as it is not prevalent in the training data. 

Second, melodies expect different levels of rhythmic granularity.
This is partly due to user annotation inconsistencies: some users annotate separate notes for every word, some for every syllable, and others only annotate when the pitch changes. Furthermore, singers often employ expressive vocal techniques, such as subtle pitch bends or grace notes. Some users annotate both the start and end of the bend, while others only annotate the core note. All the aforementioned cases serve as inconsistencies in the training and evaluation processes that compromise the performances of our model on the SheetSage-A2S dataset. 
\begin{figure}
        \centering
        \includegraphics[width=\linewidth]{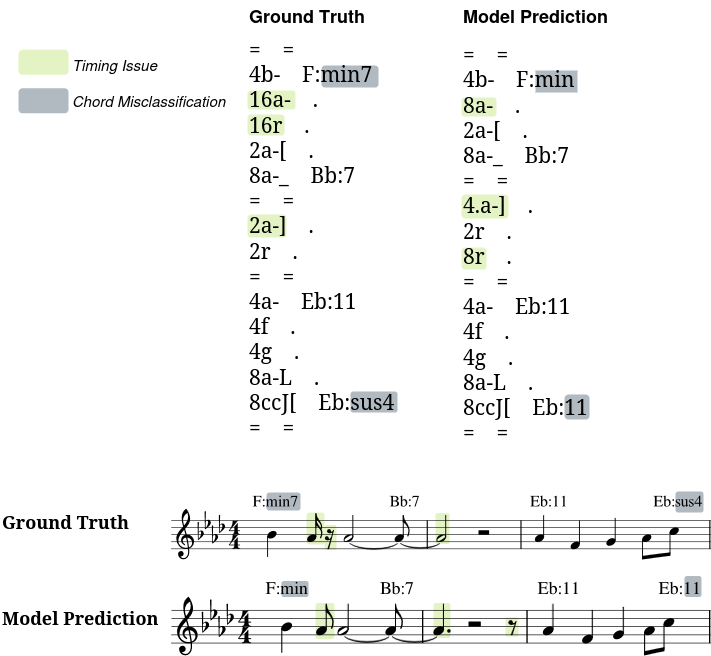}
        \caption{Measures 5 to 7 from \textit{Love Live - Bokutachi wa Hitotsu no Hikari.}\textsuperscript{12}\ It demonstrates that our best model can have trouble determining exact note-rest boundaries and non-triad chord labels. }
        \label{fig:hooktheory1}
\end{figure}

\footnotetext{\url{https://hookpad.hooktheory.com/?idOfSong=yvmrlMLVxOW} - measures 5-7}
\subsection{Qualitative analysis}

To further illustrate the model's behaviour on the SheetSage-A2S dataset, we present a qualitative analysis of the transcription results generated by our best-performing model (MuQ variant with data augmentation), along with the ground truth shown in \autoref{fig:hooktheory1}.

For chord transcription, the model typically identifies root notes correctly but has difficulty labelling chords beyond simple triads. 
Additionally, the model can sporadically label the inversion of an inverted chord as the root of the chord.
Finally, in cases where a song maintains the same root but cycles through subtly different chord qualities, the model tends to sustain the quality of its initial chord prediction without registering the subsequent changes. In \autoref{fig:hooktheory1}, for instance, the last \textit{Eb:sus4} is not identified and the \textit{Eb:11} is maintained. The disambiguation between these chords can however be subtle as both contain a perfect fourth (called 11th when an octave higher) as a characteristic tone.
Overall, 
the model successfully captures the overarching chord progression but simplifies the first F:min7 chord to a F:min.

The model's most significant difficulty lies in predicting the exact note 
durations, indicated as \textit{Timing Issue} in \autoref{fig:hooktheory1}. 
Two examples are note consolidation (merging consecutive notes) and note separation (splitting a single note). 
We observed that these issues are especially common when the same pitch is repeated multiple times in succession (common for example, in rap verse annotations). Furthermore, the model can misidentify note offsets, either cutting a note short to insert an incorrect rest, or holding a note slightly too long through a space where a 
brief rest should occur. This error is shown in \autoref{fig:hooktheory1} in green, where in the second measure the model terminates the half note prematurely by inserting an incorrect rest.

Ultimately, while this internal reshuffling causes the exact start and end times of individual notes to shift, the overall phrase length remains intact, and these timing 
deviations rarely exceed an eighth note. Musically, these offset errors are minor issues as even human transcribers might disagree on smooth offsets like those of singing vocals. Other than that, the model excels at predicting the essential musical elements, including pitches, note onsets, chord root, and the overall musical structure, which
align well with the original audio.

\section{Conclusion and Future Work}\label{sec:future}

In this paper, we first introduce the new SheetSage-A2S Dataset, a dedicated lead sheet corpus for supporting A2S research in the style of popular music. It contains 61 hours of audio for 6,066 unique songs, paired with \texttt{**kern} lead sheet annotations. We improve existing A2S frameworks by using data augmentation and MuQ \cite{zhuMuQ2025}, a pre-trained music feature extractor, to enhance generalisation and capture high-quality semantic audio features. Experiments show our A2S system significantly outperforms the existing state-of-the-art \cite{alfaro-contrerasTransformer2024} for classical music, and achieves 20.92\% SER on our SheetSage-A2S dataset for popular music, setting a competitive baseline for future research.

Looking ahead, we will expand the SheetSage-A2S dataset with instrument-performed melody clips, as current octave inference only retains clips with singing vocals. 
While this work preserves the  users' annotations with minimal pre-processing to present the SheetSage-A2S dataset in its original form, we plan to resolve the annotation variabilities and inconsistencies for the better training and evaluation of A2S models. 
We will also implement more advanced evaluation metrics, to better qualify the mistakes made by our model and their musical significance. Lastly, we plan to carry out an inter-annotator study to quantify the difference in annotation styles between annotations for our clips.

\begin{acks}
The computational resources are supported by SongShan Lake HPC Center (SSL-HPC) in Great Bay University.
\end{acks}


\newpage
\bibliographystyle{ACM-Reference-Format}
\balance
\bibliography{sample-base}


\end{document}